\documentclass[aps,prd,onecolumn,preprintnumbers,groupedaddress,showpacs,nofootinbib,amssymb]{revtex4}
\usepackage{graphicx,color}
\usepackage{amsmath}
\usepackage{amssymb}
\usepackage{amsfonts}
\usepackage{bm}
\usepackage{cancel}

\allowdisplaybreaks[4] 

\begin{document}
 
\newcommand{\be}{\begin{equation}}
\newcommand{\ee}{\end{equation}}
\newcommand{\bea}{\begin{eqnarray}}
\newcommand{\eea}{\end{eqnarray}}
\newcommand{\nn}{\nonumber \\}
\newcommand{\e}{\mathrm{e}}
\newcommand{\tr}{\mathrm{tr}\,}

\tolerance=5000

\title{Some solutions for one of the cosmological constant problems}

\author{Shin'ichi Nojiri$^{1,2}$}

\affiliation{
1. Department of Physics, Nagoya University, Nagoya
464-8602, Japan \\
2. Kobayashi-Maskawa Institute for the Origin of Particles and
the Universe, Nagoya University, Nagoya 464-8602, Japan 
}

\begin{abstract}

We propose several covariant models which may solve one of the problems 
in the cosmological constant.  
One of the model can be regarded as an extension of sequestering model.  
Other models could be regarded as extensions of the covariant formulation of the unimodular gravity. 
The contributions to the vacuum energy from the quantum corrections from the matters are absorbed 
into a redefinition of a scalar field and the quantum corrections become irrelevant to the dynamics. 
In a class of the extended unimodular gravity models, we also consider models which are 
regarded as topological field theories. 
The models can be extended and not only the vacuum energy but any quantum 
corrections to the gravitational action could become  irrelevant for the dynamics. 
We find, however, that the BRS symmetry in the topological field theories is broken spontaneously and 
therefore the models might not be consistent. 

\end{abstract}

\pacs{95.36.+x, 11.90.+t, 11.10.Ef}

\maketitle

\section{Introduction}

Now we believe that the present universe is acceleratedly expanding and the expansion is generated 
by the energy density called as dark energy, which might be a small cosmological constant. 
If the dark energy is surely a cosmological constant, there should be several problems called 
the fine-tuning problem 
or coincidence problem, whose definitions could not have been unified and different by authors. 

One problem is that if the dark energy is described by the cosmological constant $\Lambda$, 
the scale of $\Lambda$ is extremely small when we compare it with the Planck scale 
$M_\mathrm{Planck}$, which is a typical scale of the gravity, 
\be
\label{I1}
\Lambda^{1/4} \sim 10^{-3}\,\mathrm{eV} \ll M_\mathrm{Planck} 
\sim 1/\kappa
\sim 10^{19}\,\mathrm{GeV}=10^{28}\,\mathrm{eV}\, .
\ee
Here $\kappa$ is the gravitational constant. 
Another problem might be called as an anthropic principle problem. 
The energy density corresponding to the dark energy  is very small but is not so changed 
from the energy density of the present matter including the dark matter. 
If the dark energy is given by the cosmologicalonstant, this is very 
accidental and unnatural, that is, why the constant knows the energy density of the present universe? 
The above problems might be solved by considering the dynamical model of the dark energy. 

The last problem is rather old and related with the quantum correction. 
We know that in the quantum field theory, the quantum corrections from the matter 
to the energy density, 
which is called the vacuum energy $\rho_\mathrm{vacuum}$, diverge and we need 
the cutoff scale $\Lambda_\mathrm{cutoff}$, which might be the Planck scale, 
to regularize the divergence: 
\be
\label{01}
\rho_\mathrm{vacuum} 
= \frac{1}{\left( 2 \pi \right)^3}\int d^3 k \frac{1}{2} \sqrt{ k^2 + m^2 } 
\sim \Lambda_\mathrm{cutoff}^4 \, ,
\ee
which could be much larger than the observed value  $\left( 10^{-3}\, \mathrm{eV} \right)^4$ 
of the energy density in the universe.  
Even if the supersymmetry is restored in the high energy, 
the vacuum energy by the quantum corrections is given by 
$\sim \Lambda_\mathrm{cutoff}^2 \Lambda_{\cancel{\mathrm{SUSY}}}^2$. 
Here $\Lambda_{\cancel{\mathrm{SUSY}}}$ is the scale of the supersymmetry breaking.
In fact, we find
\be
\label{02}
\rho_\mathrm{vacuum} = \frac{1}{\left( 2 \pi \right)^3}\int d^3 k \frac{1}{2} 
\left( \sqrt{ k^2 + m_\mathrm{boson}^2 } 
 - \sqrt{ k^2 + m_\mathrm{fermion}^2 } \right)  
\sim \Lambda_\mathrm{cutoff}^2 \Lambda_{\cancel{\mathrm{SUSY}}}^2 \, .
\ee
Here we have assumed the scale of the supersymmetry breaking is given by the difference 
between the masses of boson and fermion: 
$\Lambda_{\cancel{\mathrm{SUSY}}}^2 
= m_\mathrm{boson}^2 - m_\mathrm{fermion}^2$ and 
we also assume $\Lambda_{\cancel{\mathrm{SUSY}}}\ll \Lambda_\mathrm{cutoff}$. 
Anyway the vacuum energy coming from the quantum corrections is very large and 
if we use the counter term in order to obtain the very small vacuum energy 
$\left(10^{-3}\, \mathrm{eV}\right)^4$, we need very very fine-tuning and 
extremely unnatural. 
This problem may tell that we could not understand quantum gravity. 

In this paper, we mainly consider the last problem only, that is, why the large quantum correction to 
the vacuum energy can be irrelevant to the dynamic, e.g., the evolution of the universe, and 
we do not discuss why the vaccum energy is so small but does not vanish. 
About the topics, see \cite{Burgess:2013ara} for example. 
Several scenarios to solve the above problem have been proposed. 
Recently an interesting mechanism called ``sequestering'' has been proposed 
 \cite{Kaloper:2013zca,Kaloper:2014dqa,Kaloper:2015jra}. 
In the next section, we briefly review the mechanism and propose an extension by using the 
Gauss-Bonnet invariant. 
Much older than the sequestering models, in order to solve the problem of the cosmological 
constant, the models called unimodular gravity have been studied 
\cite{Anderson:1971pn,Buchmuller:1988wx,Henneaux:1989zc,Unruh:1988in,
Ng:1990xz,Finkelstein:2000pg,Alvarez:2005iy,Alvarez:2006uu,Abbassi:2007bq,Ellis:2010uc,Jain:2012cw,
Singh:2012sx,Kluson:2014esa,Padilla:2014yea,Barcelo:2014mua,Barcelo:2014qva,Burger:2015kie,
Alvarez:2015sba,Jain:2012gc,Jain:2011jc,Cho:2014taa,Basak:2015swx,Gao:2014nia,Eichhorn:2015bna,
Saltas:2014cta,Nojiri:2015sfd}. 
In Section III, motivated with the unimodular gravity, we propose models,  where the contributions 
to the vacuum energy from the quantum corrections 
are absorbed into a redefinition of one of scalar fields and the contributions become irrelevant 
to the dynamics, e.g., the evolution of the universe. 
The simplest model in this class can be regarded as atopological filed theory with BRS symmetry. 
In Section IV, we investigate the poperties of the toplogical model. 
We find that the BRS symmetry in the topological field theories is broken spontaneously and 
therefore we may not be able to impose the physical state condition and therefore the models might 
not be consistent although the model has some interesting properties. 
The last section is devoted to the discussion and summary. 

\section{Sequestering Model and an Extension}

Recently an interesting mechanism to make the magnitude of the vacuum energy much smaller and consistent with the observation was proposed \cite{Kaloper:2013zca,Kaloper:2014dqa,Kaloper:2015jra}. 
In the first paper \cite{Kaloper:2013zca}, the proposed action has the following form: 
\be
\label{0Vcm1}
S = \int d^4 x \sqrt{-g } \left\{ \frac{R}{2\kappa^2} - \Lambda 
+ \e^{2\sigma} \mathcal{L}_\mathrm{matter} \left( \e^\sigma g_{\mu\nu}, \varphi \right) \right\}
 - F\left( \frac{ \Lambda \e^{-2\sigma}}{\mu^4} \right)\, .
\ee
Here $F$ is an adequate function and $\mu$ is a parameter with the dimension of mass, 
$\mathcal{L}_\mathrm{matter}$ is the Lagrangian density of matters. 
We should note that $\Lambda$ and $\varphi$ are dynamical variables which do not depend 
on the coordinate. 
In \cite{Kaloper:2013zca}, $\e^{\frac{\sigma}{2}}$ and $F$ are denoted by $\lambda$ and $\sigma$. 

The variation of $\Lambda$ gives 
\be
\label{0Vcm3}
\mu^4 F' \left( \mu^4 \Lambda \e^{-2\sigma} \right) = - \e^{2\sigma}\int d^4 x \sqrt{-g}  \, .
\ee
On the other hand, by the variation of $\sigma$, we obtain
\be
\label{0Vcm4}
\int d^4 x \sqrt{-g} \e^{-\sigma} g^{\mu\nu} 
T\left( \e^\sigma g_{\mu\nu}, \varphi \right)_{\mu\nu} 
= - 4 \mu^4 \Lambda \e^{-2\sigma} F' \left( \Lambda \e^{-2\sigma} \right) \, .
\ee
Here $T\left( \e^\sigma g_{\mu\nu}, \varphi \right)_{\mu\nu}$ is the energy-momentum tensor 
coming from the matters including the quantum corrections. 
By combining (\ref{0Vcm3}) and (\ref{0Vcm4}), we find
\be
\label{0Vcm5}
\left< \e^{-\sigma} g^{\mu\nu} T\left( \e^\sigma g_{\mu\nu}, \varphi \right)_{\mu\nu} \right> 
= 4 \Lambda \, .
\ee
Here $\left< \e^{-\sigma} g^{\mu\nu} T\left( \e^\sigma g_{\mu\nu}, \varphi \right)_{\mu\nu} 
\right>$ expresses the average of $\e^{-\sigma} g^{\mu\nu} T\left( \e^\sigma g_{\mu\nu}, \varphi 
\right)_{\mu\nu}$ with respect to the space-time,
\be
\label{03}
\left< \e^{-\sigma} g^{\rho\sigma} T\left( \e^\sigma g_{\mu\nu}, \varphi \right)_{\rho\sigma} 
\right> \equiv \frac{\int d^4 x \sqrt{-g} \e^{-\sigma} g^{\mu\nu} T\left( \e^\sigma g_{\mu\nu}, 
\varphi \right)_{\mu\nu}}{\int d^4 x \sqrt{-g}} \, .
\ee
The variation of the metric gives
\be
\label{0Vcm6}
0= - \frac{1}{2\kappa^2}\left( R_{\mu\nu} - \frac{1}{2} R g_{\mu\nu}\right) 
 - \frac{1}{2} \Lambda g_{\mu\nu}
+ \frac{1}{2}T\left( \e^\sigma g_{\mu\nu}, \varphi \right)_{\mu\nu} \, .
\ee
By using (\ref{0Vcm5}), we may rewrite (\ref{0Vcm6}) as follows,
\be
\label{0Vcm7}
0= - \frac{1}{2\kappa^2}\left( R_{\mu\nu} - \frac{1}{2} R g_{\mu\nu}\right) 
 - \frac{1}{8} \left< \e^{-\sigma} g^{\rho\sigma} T\left( \e^\sigma g_{\mu\nu}, \varphi \right)_{\rho\sigma} \right> g_{\mu\nu}
+ \frac{1}{2}T\left( \e^\sigma g_{\mu\nu}, \varphi \right)_{\mu\nu}\, .
\ee
In the combination $ - \frac{1}{8} \left< \e^{-\sigma} g^{\rho\sigma} 
T\left( \e^\sigma g_{\mu\nu}, \varphi \right)_{\rho\sigma} \right> g_{\mu\nu}
+ \frac{1}{2}T\left( \e^\sigma g_{\mu\nu}, \varphi \right)_{\mu\nu}$, 
the large quantum correction to the vacuum energy is canceled. 

After \cite{Kaloper:2013zca}, several extension of the model have been considered 
\cite{Kaloper:2014dqa,Kaloper:2015jra}. 
In \cite{Kaloper:2015jra}, instead of the global variables $\Lambda$ and $\sigma$, 4-form field are 
introduced and the model can be written in a totally local form. 
Even in the model \cite{Kaloper:2015jra}, the constraint corresponding to (\ref{03}) is global and there 
might appear any problem related with the causality. 

Motivated by the papers \cite{Kaloper:2013zca,Kaloper:2014dqa,Kaloper:2015jra}, 
we consider the following action. 
\be
\label{Vcm1}
S = \int d^4 x \sqrt{-g } \left\{ \frac{R}{2\kappa^2} - \Lambda(x) + \e^{2\sigma(x)} \mathcal{L}_\mathrm{matter} \left( \e^\sigma g_{\mu\nu}, \varphi \right) 
 - f\left( \mu^4 \Lambda(x) \e^{-2\sigma(x)} \right)\mathcal{G} \right\}\, .
\ee
Here $f$ is a function which may be determined later. 
We should note that now $\Lambda(x)$ and $\sigma (x)$ are not global variables but dynamical variables. 
Instead of the 4-form field in \cite{Kaloper:2015jra}, we introduce the Gauss-Bonnet invariant 
$\mathcal{G}$ defined by 
\be
\label{Vcm2}
\mathcal{G} \equiv R^2 - 4 R_{\mu\nu} R^{\mu\nu} + R_{\mu\nu\rho\sigma} 
R^{\mu\nu\rho\sigma} \, .
\ee
The Gauss-Bonnet invariant $\mathcal{G}$, of course, depend on the metric $g_{\mu\nu}$ but because 
$\mathcal{G}$ is a total derivative, when $\Lambda(x) \e^{-2\sigma}$ is a constant, the term including 
$\mathcal{G}$ does not contribute to the equation given by the variation of the metric. 

The variation of $\Lambda$ gives 
\be
\label{Vcm3}
\mu^4 f' \left( \mu^4 \Lambda(x) \e^{-2\sigma} \right) \mathcal{G}= - \e^{2\sigma} \, .
\ee
On the other hand, by the variation of $\sigma$, we obtain
\be
\label{Vcm4}
\e^{-\sigma} g^{\mu\nu} T\left( \e^\sigma g_{\mu\nu}, \varphi \right)_{\mu\nu} 
=   - 4 \mu^4 \Lambda(x) \e^{-2\sigma} f' \left( \Lambda(x) \e^{-2\sigma} \right) \mathcal{G}\, .
\ee
Here $T\left( \e^\sigma g_{\mu\nu}, \varphi \right)_{\mu\nu}$ is the energy-momentum tensor coming 
from the matters including the quantum corrections. 
By combining (\ref{Vcm3}) and (\ref{Vcm4}), we find
\be
\label{Vcm5}
\e^{-\sigma} g^{\mu\nu} T\left( \e^\sigma g_{\mu\nu}, \varphi \right)_{\mu\nu} 
=   4 \Lambda(x) \, .
\ee
The variation of the metric gives
\begin{align}
\label{Vcm6}
0=& - \frac{1}{2\kappa^2}\left( R_{\mu\nu} - \frac{1}{2} R g_{\mu\nu}\right) 
 - \frac{1}{2} \Lambda (x) g_{\mu\nu}
+ \frac{1}{2}T\left( \e^\sigma g_{\mu\nu}, \varphi \right)_{\mu\nu}
 - 2 \left[ \nabla_\mu \nabla_\nu \left( 
f \left( \mu^4 \Lambda(x) \e^{-2\sigma} \right) 
 \right) \right] R \nn
& + 4 \left[ \nabla_\mu \nabla_\rho \left( 
f \left( \mu^4 \Lambda(x) \e^{-2\sigma} \right) 
\right) \right] R_\nu^{\ \rho} 
+ 4 \left[ \nabla_\nu \nabla_\rho \left( 
f \left( \mu^4 \Lambda(x) \e^{-2\sigma} \right) 
\right) \right] R_\mu^{\ \rho} \nn
& - \left[ \nabla_\rho \nabla^\rho \left( 
f \left( \mu^4 \Lambda(x) \e^{-2\sigma} \right) 
\right) \right] \left(4 R_{\mu\nu} - 2 R g_{\mu\nu} \right) 
 -4 \left[ \nabla_\rho \nabla_\sigma \left( 
f \left( \mu^4 \Lambda(x) \e^{-2\sigma} \right) 
\right)\right] \left(R^{\rho\sigma} g_{\mu\nu} 
 - R_{\mu\ \nu\ }^{\ \rho\ \sigma} \right)\ .
\end{align}
By defining $\Phi(x) \equiv f \left( \mu^4 \Lambda(x) \e^{-2\sigma} \right)$ and by using (\ref{Vcm5}), 
we may rewrite (\ref{Vcm7}) as follows,
\begin{align}
\label{Vcm7}
0=& - \frac{1}{2\kappa^2}\left( R_{\mu\nu} - \frac{1}{2} R g_{\mu\nu}\right) 
 - \frac{1}{8} \e^{-\sigma} g^{\rho\sigma} 
T\left( \e^\sigma g_{\mu\nu}, \varphi \right)_{\rho\sigma} g_{\mu\nu}
+ \frac{1}{2}T\left( \e^\sigma g_{\mu\nu}, \varphi \right)_{\mu\nu}
 - 2 \left[ \nabla_\mu \nabla_\nu \Phi(x) \right] R \nn
& + 4 \left[ \nabla_\mu \nabla_\rho \Phi \right] R_\nu^{\ \rho} 
+ 4 \left[ \nabla_\nu \nabla_\rho \Phi \right] R_\mu^{\ \rho} 
 - \left[ \nabla_\rho \nabla^\rho \Phi \right] \left(4 R_{\mu\nu} - 2 R g_{\mu\nu} \right) 
 -4 \left[ \nabla_\rho \nabla_\sigma \Phi \right] \left(R^{\rho\sigma} g_{\mu\nu} 
 - R_{\mu\ \nu\ }^{\ \rho\ \sigma} \right)\ .
\end{align}
In the combination $ - \frac{1}{8} \e^{-\sigma} g^{\rho\sigma} T\left( \e^\sigma g_{\mu\nu}, \varphi 
\right)_{\rho\sigma} g_{\mu\nu} 
+ \frac{1}{2}T\left( \e^\sigma g_{\mu\nu}, \varphi \right)_{\mu\nu}$, 
the large quantum correction to the vacuum energy is canceled locally. 
We may now estimate the magnitude of the Gauss-Bonnet term. 
We now assume the function $f(x) \propto x^\alpha$ with a constant $\alpha$. 
By assuming that the cut off scale is given by the Planck scale, we find 
\be
\label{Vcm8}
\e^{-\sigma} g^{\mu\nu} T\left( \e^\sigma g_{\mu\nu}, \varphi \right)_{\mu\nu} 
=   4 \Lambda(x) \sim \mu^{-4} \sim \left( 10^{28}\, \mathrm{eV} \right)^4\, .
\ee
Here we have assumed the scale of $\mu$ is that of the Planck scale, $\mu \sim 10^{28}\, \mathrm{eV}$. 
In the present universe, we find $\mathcal{G} \sim \left( 10^{-33}\, \mathrm{eV} \right)^4$. 
Then by using (\ref{Vcm3}), we obtain 
\be
\label{Vcm9}
f \left( \mu^4 \Lambda(x) \e^{-2\sigma} \right) \sim \e^{- 2\alpha \sigma} \sim 10^{244}\, ,
\ee
which does not depend on $\alpha$ and very large. 
In the physical frame where the metric is given by 
$g_{\mu\nu}^\mathrm{phys} \equiv \e^\sigma g_{\mu\nu}$, however, we find 
\be
\label{Vcm10}
\int d^4 x \sqrt{-g }  f\left( \mu^4 \Lambda(x) \e^{-2\sigma(x)} \right)\mathcal{G} 
= \int d^4 x \sqrt{-g^\mathrm{phys} } \e^{-2\sigma} f\left( \mu^4 \Lambda(x) 
\e^{-2\sigma(x)} \right)\left( \mathcal{G}^\mathrm{phys} + \cdots \right)\, .
\ee
Here $\mathcal{G}^\mathrm{phys}$ is the Gauss-Bonnet invariant given by 
$g_{\mu\nu}^\mathrm{phys}$ and ``$\cdots$'' expresses the terms including the derivatives of 
$\sigma$. 
Then we find 
\be
\label{Vcm11}
\e^{-2\sigma} f\left( \mu^4 \Lambda(x) \e^{-2\sigma(x)} \right)
= \e^{244 \left( 1 + \frac{1}{\alpha} \right)}\, .
\ee
Then if choose that $\alpha$ to be negative and the absolute value of $\alpha$ to be  small, 
$\e^{-2\sigma} f\left( \mu^4 \Lambda(x) \e^{-2\sigma(x)} \right)$ can be arbitrarily small and the 
terms from the Gauss-Bonnet term in (\ref{Vcm7}) can be negligible. 
On the other hand if we tune the parameter $\alpha$, the Gauss-Bonnet term 
may explain the accelerating expansion of the present universe.  

\section{Generalization of Unimodular Gravity}

Models older than the sequestering models in the last section, which is  called unimodular gravity, 
have also properties similar to the sequestering models 
\cite{Anderson:1971pn,Buchmuller:1988wx,Henneaux:1989zc,Unruh:1988in,
Ng:1990xz,Finkelstein:2000pg,Alvarez:2005iy,Alvarez:2006uu,Abbassi:2007bq,Ellis:2010uc,Jain:2012cw,
Singh:2012sx,Kluson:2014esa,Padilla:2014yea,Barcelo:2014mua,Barcelo:2014qva,Burger:2015kie,
Alvarez:2015sba,Jain:2012gc,Jain:2011jc,Cho:2014taa,Basak:2015swx,Gao:2014nia,Eichhorn:2015bna,
Saltas:2014cta,Nojiri:2015sfd}.\footnote{
There are several other scenario to solve the cosmological constant problems,e.g., 
\cite{Batra:2008cc,Shaw:2010pq,Barrow:2010xt,Carballo-Rubio:2015kaa}. 
} 
The model in this section can be regarded as an extension of the unimodular gravity in the covariant 
formulation \cite{Henneaux:1989zc,Buchmuller:1988yn} but the model in this section 
is more general. 

In the unimodular gravity, the determinant of the metric is constrained to be unity,
\be
\label{Uni1}
\sqrt{-g}=1\, ,
\ee
which is called a unimodular constraint. 
In the Lagrangian formalism, the constraint can be realized by using the Lagrange multiplier field $\lambda$ 
(see \cite{Buchmuller:1988yn,Henneaux:1989zc}, for example) 
as follows:
\be
\label{UniCC0}
S = \int d^4 x \left\{ \sqrt{-g} \left( \mathcal{L}_\mathrm{gravity} - \lambda \right) 
+ \lambda \right\} + S_\mathrm{matter} \, .
\ee
Here $S_\mathrm{matter}$ is the action of matters and $\mathcal{L}_\mathrm{gravity}$ is 
the Lagrangian density of arbitrary gravity models. 
By the variation of $\lambda$, we obtain the unimodular constraint (\ref{Uni1}). 
We may divide the gravity Lagrangian density $\mathcal{L}_\mathrm{gravity}$ into the sum of 
the cosmological constant $\Lambda$ and other part $\mathcal{L}_\mathrm{gravity}^{(0)}$ as follows: 
\be
\label{UniCC1}
\mathcal{L}_\mathrm{gravity} = \mathcal{L}_\mathrm{gravity}^{(0)} - \Lambda \, .
\ee
We may also redefine the Lagrange multiplier field $\lambda$ by $\lambda \to \lambda - \Lambda$. 
Then the action (\ref{UniCC0}) can be rewritten as 
\be
\label{UniCC2}
S = \int d^4 x \left\{ \sqrt{-g} \left( \mathcal{L}_\mathrm{gravity}^{(0)} - \lambda \right) 
+ \lambda \right\} + S_\mathrm{matter} + \Lambda \int d^4 x \, .
\ee
Because the last term $\Lambda \int d^4 x$ does not depend on any dynamical variable, we may drop 
the last term. 
Therefore the obtained action (\ref{UniCC2}) does not include the cosmological constant.  
This tells that the cosmological constant $\Lambda$ does not affect the dynamics even in the action 
(\ref{UniCC1}).  
We should note that the cosmological constant may include the large quantum corrections from matters 
to the vacuum energy. 
Because the cosmological constant $\Lambda$ does not affect the dynamics, the large quantum 
corrections can be tuned to vanish. 

Due to the unimodular constraint (\ref{Uni1}), the unimodular gravity does not have full covariance.
The covariant formulation of the unimodular Einstein gravity has been proposed 
in \cite{Henneaux:1989zc,Buchmuller:1988yn}, where the action is given by 
\be
\label{LUF1}
S = \int d^4 x \left\{ \sqrt{-g} \left( \mathcal{L}_\mathrm{gravity} 
 - \lambda \right) + \lambda 
\epsilon^{\mu\nu\rho\sigma} \partial_\mu a_{\nu\rho\sigma} \right\}
+ S_\mathrm{matter} \left( g_{\mu\nu}, \Psi \right)\, .
\ee
Here $a_{\nu\rho\sigma}$ is the three-form field.
The variation over $a_{\nu\rho\sigma}$ gives the equation
$0 = \partial_\mu \lambda$, that is, $\lambda$ is a constant.
On the other hand, the variation over $\lambda$ gives
\be
\label{LLUF19}
\sqrt{-g} = \epsilon^{\mu\nu\rho\sigma} \partial_\mu a_{\nu\rho\sigma}\, ,
\ee
instead of the unimodular constraint (\ref{Uni1}).
Because Eq.~(\ref{LLUF19}) can be solved with respect to $a_{\mu\nu\rho}$, 
there is no constraint on the metric $g_{\mu\nu}$.
If we divide the gravity Lagrangian density $\mathcal{L}_\mathrm{gravity}$ into the sum of 
the cosmological constant $\Lambda$ and other part $\mathcal{L}_\mathrm{gravity}^{(0)}$ as in 
(\ref{UniCC1}) and redefining the Lagrange multiplier field $\lambda$ by 
$\lambda \to \lambda - \Lambda$, we can rewrite the action (\ref{CCC2}) as follows, 
\be
\label{LUF1B}
S = \int d^4 x \left\{ \sqrt{-g} \left( \mathcal{L}_\mathrm{gravity}^{(0)} 
 - \lambda \right) + \lambda 
\epsilon^{\mu\nu\rho\sigma} \partial_\mu a_{\nu\rho\sigma} \right\} 
+ \Lambda \int d^4 x  
\epsilon^{\mu\nu\rho\sigma} \partial_\mu a_{\nu\rho\sigma} 
+ S_\mathrm{matter} \left( g_{\mu\nu}, \Psi \right)\, .
\ee
Because the term $\int d^4 x  \epsilon^{\mu\nu\rho\sigma} \partial_\mu a_{\nu\rho\sigma} $ 
is a total derivative, the term does not give any dynamical contribution and can be droped. 
Then the obtained action does not include the cosmological constant, which tells that the cosmological 
constant does not contribute to any dynamics.  

We may generalize the covariant formulation in the unimodular gravity 
and we now propose the following model, 
\be
\label{CCC2}
S = \int d^4 x \sqrt{-g} \left\{ \mathcal{L}_\mathrm{gravity}
 - \lambda \left( 1 - \frac{1}{\mu^4}\nabla_\mu J^\mu  \right) \right\} 
+ S_\mathrm{matter} \, .
\ee
Here $J^\mu$ is a general vector quantity, 
and $\nabla_\mu$ is a covariant derivative with respect to the vector field. 
Then by dividing the gravity Lagrangian density $\mathcal{L}_\mathrm{gravity}$ into the sum of 
the cosmological constant $\Lambda$ and other part $\mathcal{L}_\mathrm{gravity}^{(0)}$ as in 
(\ref{UniCC1}) and redefining the Lagrange multiplier field $\lambda$ by 
$\lambda \to \lambda - \Lambda$, again, we can rewrite the action (\ref{CCC2}) as follows, 
\be
\label{CCC2b}
S = \int d^4 x \sqrt{-g} \left\{ \mathcal{L}_\mathrm{gravity}^{(0)}
 - \lambda \left( 1 - \frac{1}{\mu^4}\nabla_\mu J^\mu  \right) \right\} 
+ S_\mathrm{matter} 
 -  \frac{\Lambda}{\mu^4} \int d^4 x \sqrt{-g} \nabla_\mu J^\mu \, .
\ee
Because the integrand in the last term is total derivative, again, we may drop the last term. 
We may consider the following action instead of (\ref{CCC2}), 
\be
\label{CCC2b}
S = \int d^4 x \sqrt{-g} \left\{ \mathcal{L}_\mathrm{gravity}
 - \lambda - \frac{1}{\mu^4} \nabla_\mu \lambda J^\mu \right\} 
+ S_\mathrm{matter} \, .
\ee
The difference of the action (\ref{CCC2b}) from (\ref{CCC2}) is the total derivative but 
by the redefinition of the Lagrange multiplier field $\lambda$ by 
$\lambda \to \lambda - \Lambda$, there does not appear the total derivative term but the 
cosmological constant $\Lambda$ can be locally absorbed. 

We may choose $\nabla_\mu J^\mu$ to be a topological invariant like the Gauss-Bonnet invariant 
$\mathcal{G}$ in (\ref{Vcm2}), 
$I \equiv \frac{1}{4!} \epsilon^{\mu\nu\rho\sigma} F_{\mu\nu} F_{\rho\sigma}$ for 
abelian gauge theory, or instanton density 
$I \equiv \frac{1}{4!} \epsilon^{\mu\nu\rho\sigma} \tr F_{\mu\nu} F_{\rho\sigma}$ for 
non-abelian abelian gauge theory. 

There are several variations in the action (\ref{CCC2}) by the choice of $J^\mu$. 
For example, instead of the topological invariants, we may include a complex scalar field $\phi$ 
and consider the following action, 
\be
\label{CCC3}
S = \int d^4 x \sqrt{-g} \left\{ \mathcal{L}_\mathrm{gravity} 
 - \frac{2} \partial_\mu \phi^* \partial^\mu \phi 
 - \lambda \left( 1 - \frac{i}{\mu^4}\nabla_\mu \left( \phi^* \partial^\mu \phi 
 - \left(\partial^\mu \phi^* \right) \phi \right) \right) \right\} 
+ S_\mathrm{matter} \, .
\ee
More simple model which may be regarded with a toplogical field theory is given in next section.  

We may further propose a new class of models, which may also solve the problem of the 
vacuum energy and the action is given by
\be
\label{NewCCC}
S = \int d^4 x \sqrt{-g} \left\{ \mathcal{L}_\mathrm{gravity}  - \lambda  
+ \mathcal{L}_\lambda \left( \partial_\mu, \partial_\mu \lambda, \varphi_i \right) \right\}
+ S_\mathrm{matter} \, ,
\ee
where $\mathcal{L}_\lambda \left( \partial_\mu, \partial_\mu \lambda, \varphi_i \right)$ is 
the Lagrangian density including the derivatives of $\lambda$ and other
fields $\varphi_i$, but not including $\lambda$ without derivative.
Hence, if we divide the  Lagrangian density $\mathcal{L}_\mathrm{gravity}$
into the sum of the cosmological constant $\Lambda$ and other part
$\mathcal{L}_\mathrm{gravity}^{(0)}$, the cosmological constant can be
absorbed into the redefinition of the Lagrange multiplier field
$\lambda$, $\lambda \to \lambda - \Lambda$. 
The Lagrangian density 
$\mathcal{L}_\lambda \left( \partial_\mu, \partial_\mu \lambda, \varphi_i \right)$ 
can be that of the massless scalar field, 
\be
\label{A1}
\mathcal{L}_\lambda \left( \partial_\mu, \partial_\mu \lambda, \varphi_i\right)
= - \frac{1}{2} g^{\mu\nu} \partial_\mu \lambda \partial_\nu \lambda \, ,
\ee
or that of the $k$-essence 
\cite{ArmendarizPicon:1999rj,Garriga:1999vw,ArmendarizPicon:2000ah,Chiba:1997ej}, 
\be
\label{A2}
\mathcal{L}_\lambda \left( \partial_\mu, \partial_\mu \lambda, \varphi_i\right)
= \mathcal{L} \left( - \frac{1}{2} g^{\mu\nu} \partial_\mu \lambda \partial_\nu \lambda
\right) \, .
\ee
Here $\mathcal{L} \left( - \frac{1}{2} g^{\mu\nu} \partial_\mu \lambda \partial_\nu \lambda
\right)$ is a general function of $- \frac{1}{2} g^{\mu\nu} \partial_\mu \lambda \partial_\nu \lambda$. 
Furthermore the Lagrangian density can be that of the Galileon model 
\cite{Nicolis:2008in}. 
Then these models can describe more realistic and complex evolutions of the universe. 

\section{Topological Model}

In a class of models in (\ref{CCC2}), we may aconsider a simpler, maybe simplest, model 
by using a real scalar field $\varphi$ as follows:\footnote{
This action itself has appeared in \cite{Shlaer:2014gna} for other purpose. 
As we discuss, the model in (\ref{CCC4}) includes ghost and therefore the model is 
inconsistent but we propose a new model (\ref{CCC7}) in order to avoid the problem of the ghost. 
}
\begin{align}
\label{CCC4}
S =& \int d^4 x \sqrt{-g} \left\{ \mathcal{L}_\mathrm{gravity} 
 - \lambda \left( 1 + \frac{1}{\mu^3}\nabla_\mu \partial^\mu \varphi \right) \right\} 
+ S_\mathrm{matter} \nn
=& \int d^4 x \sqrt{-g} \left\{ \mathcal{L}_\mathrm{gravity} 
 - \lambda + \frac{1}{\mu^3} \partial_\mu \lambda \partial^\mu \varphi \right\} 
+ S_\mathrm{matter} \, .
\end{align}
The term $\frac{1}{\mu^3} \partial_\mu \lambda \partial^\mu \varphi$ in the action (\ref{CCC4}) 
tells that this model may include a ghost. 
In fact if we redefine the scalar fields $\varphi$ and $\lambda$ as follows, 
\be
\label{CCC5}
\varphi = \frac{1}{\sqrt{2}}\left( \eta + \xi \right) \, , \quad
 \lambda = \frac{\mu^3}{\sqrt{2}} \left( \eta - \xi \right) \, ,
\ee
the action can be rewritten as
\be
\label{CCC6} 
S = \int d^4 x \sqrt{-g} \left\{ \mathcal{L}_\mathrm{gravity} 
 - \frac{1}{2} \partial_\mu \xi \partial^\mu \xi 
+ \frac{1}{2} \partial_\mu \eta \partial^\mu \eta + \frac{\mu^3}{\sqrt{2}} 
\left( \eta - \xi \right) \right\} + S_\mathrm{matter} \, .
\ee
The kinetic term of $\eta$ tells that the scalar field $\eta$ generates the negative norm state and 
therefore $\eta$ is a ghost. 
This problem of the ghost may be avoided by introducing the fermionic (Grassmann odd) ghosts 
$b$ and $c$, 
\be
\label{CCC7} 
S' = \int d^4 x \sqrt{-g} \left\{ \mathcal{L}_\mathrm{gravity} 
 - \lambda + \frac{1}{\mu^3} \partial_\mu \lambda \partial^\mu \varphi  
 - \partial_\mu b \partial^\mu c \right\} + S_\mathrm{matter} \, .
\ee
The action is invariant under the BRS transformation \cite{Becchi:1975nq},
\be
\label{CCC8}
\delta \lambda = \delta c = 0\, , \quad 
\delta \varphi = \epsilon c \, , \quad \delta b = \frac{1}{\mu^3} \epsilon \lambda \, .
\ee
Here $\epsilon$ is a fermionic parameter. 
Then by defining the physical states as the states invariant under the BRS transformation, 
the negative norm states could be removed as in the gauge theory \cite{Kugo:1977zq,Kugo:1979gm}. 
If we assign the ghost number $1$ for $c$ and $-1$ for $b$ and $\epsilon$, the ghost number is 
also conserved. 
We can identify $\lambda$, $\varphi$, $b$, and $c$ with a quartet in Kugo-Ojima's quartet mechanism 
in the gauge theory \cite{Kugo:1977zq,Kugo:1979gm}. 

In the action (\ref{CCC8}) the Lagrangian density,
\be
\label{SCCP1}
\mathcal{L} =  - \lambda + \frac{1}{\mu^3} \partial_\mu \lambda \partial^\mu \varphi  
 - \partial_\mu b \partial^\mu c \, ,
\ee
can be regarded as the Lagrangian density of a topological field theory \cite{Witten:1988ze}, 
where the Lagrangian density is BRS exact, that is, given by the BRS transformation of some quantity. 
We start with the field theory of $\varphi$ but the Lagrangian density vanishes $\mathcal{L}_\varphi=0$. 
Because the Lagrangian density vanishes, under any transformation of $\varphi$, the Lagrangian density 
is trivially invariant. 
Then this theory can be regarded as a gauge theory. 
In order to fix the gauge, we impose the following gauge condition, 
\be
\label{CCC9}
1 + \frac{1}{\mu^3} \nabla_\mu \partial^\mu \varphi = 0\, .
\ee
Then the gauge-fixing Lagrangian \cite{Kugo:1981hm} is given by the BRS transformation (\ref{CCC8}) of 
$- b \left( 1 + \frac{1}{\mu^3} \nabla_\mu \partial^\mu \varphi \right)$.
In fact, we find 
\be
\label{SCCP2}
\delta \left(- b \left( 1 + \frac{1}{\mu^3} \nabla_\mu \partial^\mu \varphi \right) \right)
= \epsilon \left( - \lambda  \left( 1 + \frac{1}{\mu^3} \nabla_\mu \partial^\mu \varphi \right) 
+ b \nabla_\mu \partial^\mu c \right) 
= \epsilon \left( \mathcal{L} + \left(\mbox{total derivative terms}\right) \right)\, .
\ee
Therefore the Lagrangian density (\ref{SCCP1}) is surely BRS exact up to total derivative and 
the theory described by the Lagrangian density could a toplogical field theory. 

Eq.~(\ref{CCC8}) tells that $\lambda$ corresponds to the Nakanishi-Lautrup field 
\cite{Nakanishi:1966zz,Nakanishi:1973fu,Lautrup:1967zz} and as we find from Eq.~(\ref{CCC8}),
$\lambda$ is BRS exact, which tells that the vacuum expectation value of $\lambda$ should vanish. 
If the vacuum expectation value of $\lambda$ does not vanish, the BRS symmetry is spontaneously broken 
and we may not be able to consistently impose the physical state condition.  
This might be true only for the oscillating mode but the zero-mode could not always need to vanish.  
In the formulation by using the Nakanishi-Lautrup field, the physical states are the states 
annihilated by the positive frequency part of the Nakanishi-Lautrup field, which does not include 
the zero frequency or non-oscillating mode. 
On the other hand, the BRS charge $Q_\mathrm{BRS}$ corresponding to 
the BRS transformation (\ref{CCC8}) is given by
\be
\label{BRS}
Q_\mathrm{BRS} = \frac{1}{\mu^3} \int d^3 x \sqrt{-g} \left( 
\partial^0 \lambda c - \lambda \partial^0 c \right)  \, .
\ee
Because $\lambda$ and $c$ satisify the equations 
$\nabla_\mu \partial^\mu \lambda = \nabla_\mu \partial^\mu c = 0$, 
if $\lambda$ and $c$ are constant, we may obtain $\partial^0 \lambda = \partial^ c = 0$ if we impose 
boundary or initial conditions where $\lambda$ and $c$ are finite in the infinite future or past. 
If it could be true, the BRS charge $Q_\mathrm{BRS}$ does not include the constant mode of $\lambda$ 
nor $c$ and therefore $Q_\mathrm{BRS}$ anti-commutes with the constant mode of $b$ and 
the constant mode of $\lambda$ could not be BRS exact and could be able to have non-vanishing expectation 
value in the vacuum. 

We should also note that the action (\ref{CCC7}) is also invariant under the following deformation 
of the BRS transformation (\ref{CCC8}),
\be
\label{CCC8B}
\delta \lambda = \delta c = 0\, , \quad 
\delta \varphi = \epsilon c \, , \quad 
\delta b = \frac{1}{\mu^3} \epsilon \left( \lambda + \lambda_0 \right)\, .
\ee
Here $\lambda_0$ is an arbitrary constant. 
Although it might not be natural, if we can choose $\lambda_0$ to be the cosmological constant 
$\Lambda$ as $\lambda_0=\Lambda$, the quantity $\lambda + \Lambda$ becomes BRS exact and the 
expectation value by any physical states vanishes and therefore $\lambda$ exactly cancells the cosmological 
constant $\Lambda$. 

We have considered the quantum corrections to the vacuum energy coming from the quantum 
corrections of matter. 
We should note, however, that the vacuum energy is not only the quantum corrections but by the 
quantum corrections from the matter, the following terms are generated, 
\be
\label{A3}
\mathcal{L}_\mathrm{qc} = \alpha R + \beta R^2 + \gamma R_{\mu\nu} R^{\mu\nu} 
+ \delta R_{\mu\nu\rho\sigma} R^{\mu\nu\rho\sigma} \, .
\ee
Here the coefficient $\alpha$ diverges quadrutically and $\beta$, $\gamma$, and $\delta$ diverge 
logarithmically without the cut-off scale. 
Then a generation of the (\ref{SCCP1}) is given by 
\begin{align}
\label{A4}
\mathcal{L} =&   - \Lambda - \lambda_{(\Lambda)} 
+ \left( \alpha  + \lambda_{(\alpha)} \right)R 
+ \left( \beta + \lambda_{(\beta)} \right) R^2 
+ \left( \gamma + \lambda_{(\gamma)} \right) R_{\mu\nu} R^{\mu\nu} 
+ \left( \delta + \lambda_{(\delta)} \right) 
R_{\mu\nu\rho\sigma} R^{\mu\nu\rho\sigma} \nn
& + \frac{1}{\mu^3} \partial_\mu \lambda_{(\Lambda)} \partial^\mu \varphi_{(\Lambda)}  
 - \partial_\mu b_{(\Lambda)} \partial^\mu c_{(\Lambda)} 
+ \frac{1}{\mu} \partial_\mu \lambda_{(\alpha)} \partial^\mu \varphi_{(\alpha)}  
 - \partial_\mu b_{(\alpha)} \partial^\mu c_{(\alpha)} \nn
& + \mu \partial_\mu \lambda_{(\beta)} \partial^\mu \varphi_{(\beta)}  
 - \partial_\mu b_{(\beta)} \partial^\mu c_{(\beta)} 
+ \mu \partial_\mu \lambda_{(\gamma)} \partial^\mu \varphi_{(\gamma)}  
 - \partial_\mu b_{(\gamma)} \partial^\mu c_{(\gamma)} 
+ \mu \partial_\mu \lambda_{(\delta)} \partial^\mu \varphi_{(\delta)}  
 - \partial_\mu b_{(\delta)} \partial^\mu c_{(\delta)} \, .
\end{align}
The coefficients $\Lambda$, $\alpha$, $\beta$, $\gamma$, and $\delta$ may include 
the divergences due to the quantum corrections from the matters 
but if we consider the redefinitions of the parameters, 
$\lambda_{(0})$, $\lambda_{(\alpha)}$, $\lambda_{(\beta)}$, $\lambda_{(\gamma)}$, 
and $\lambda_{(\delta)}$ by
\be
\label{A5}
\lambda_{(\Lambda)} \to \lambda_{(\lambda)} - \Lambda\, , \quad 
\lambda_{(\alpha)} \to \lambda_{(\alpha)}  - \alpha\, , \quad 
\lambda_{(\beta)} \to \lambda_{(\beta)} - \beta\, , \quad 
\lambda_{(\gamma)} \to \lambda_{(\gamma)} - \gamma\, , \quad 
\lambda_{(\delta)} \to \lambda_{(\delta)} - \delta \, ,
\ee
the Lagrangian density (\ref{A4}) can be rewritten as 
\begin{align}
\label{A6}
\mathcal{L} =&   - \lambda_{(\Lambda)} 
+ \lambda_{(\alpha)} R + \lambda_{(\beta)} R^2 
+ \lambda_{(\gamma)} R_{\mu\nu} R^{\mu\nu} 
+ \lambda_{(\delta)} R_{\mu\nu\rho\sigma} R^{\mu\nu\rho\sigma} \nn
& + \frac{1}{\mu^3} \partial_\mu \lambda_{(\Lambda)} \partial^\mu \varphi_{(\Lambda)}  
 - \partial_\mu b_{(\Lambda)} \partial^\mu c_{(\Lambda)} 
+ \frac{1}{\mu} \partial_\mu \lambda_{(\alpha)} \partial^\mu \varphi_{(\alpha)}  
 - \partial_\mu b_{(\alpha)} \partial^\mu c_{(\alpha)} \nn
& + \mu \partial_\mu \lambda_{(\beta)} \partial^\mu \varphi_{(\beta)}  
 - \partial_\mu b_{(\beta)} \partial^\mu c_{(\beta)} 
+ \mu \partial_\mu \lambda_{(\gamma)} \partial^\mu \varphi_{(\gamma)}  
 - \partial_\mu b_{(\gamma)} \partial^\mu c_{(\gamma)} 
+ \mu \partial_\mu \lambda_{(\delta)} \partial^\mu \varphi_{(\delta)}  
 - \partial_\mu b_{(\delta)} \partial^\mu c_{(\delta)} \, .
\end{align}
Then the divergences can be absorbed into the redefinition of $\lambda_{(i)}$, 
$\left(i=\Lambda,\alpha,\beta,\gamma,\delta\right)$ and therefore the divergences might become 
irrelevant for the dynamics. 
The Lagrangian density (\ref{A6}) is invariant under the BRS transformation 
\be
\label{A7}
\delta \lambda_{(i)} = \delta c_{(i)} = 0\, , \quad 
\delta \varphi_{(i)}i = \epsilon c \, , \quad 
\delta b_{(i)} = \frac{1}{\mu^k} \epsilon \lambda_{(i)} \, , 
\quad \left(i=\Lambda,\alpha,\beta,\gamma,\delta\right)\, ,
\ee
and $k=3$ for $i=\Lambda$, $k=1$ for $i=\alpha$, and $k=-1$ for $i=\beta,\gamma,\delta$.  
The Lagrangian density (\ref{A6}) is also BRS exact, 
\be
\label{A8}
\delta \left( \sum_{i=0,\alpha,\beta,\gamma,\delta} 
\left(- b_{(i)} \left( 1 + \frac{1}{\mu^k} \nabla_\mu \partial^\mu \varphi_{(i)} 
\right) \right) \right)
= \epsilon \left( \mathcal{L} + \left(\mbox{total derivative terms}\right) \right)\, .
\ee

If we include the quantum corrections from the graviton, there appear infinite numbers of 
quantum corrections which diverge. 
Letting $\mathcal{O}_i$ be possible gravitational operators, a further generalization of the 
lagrangian density (\ref{A6}) could be given by
\be
\label{A11}
\mathcal{L} = \sum_i \left( \lambda_{(i)} \mathcal{O}_{(i)} 
+ \frac{1}{\mu^{k_{(i)}}} \partial_\mu \lambda_{(i)} \partial^\mu \varphi_{(i)}  
 - \partial_\mu b_{(i)} \partial^\mu c_{(i)} \right) \, .
\ee
In the Lagrangian density (\ref{A11}), all the divergence can be absorbed into the 
redefinition of $\lambda_i$ and become irrelevant for the dynamics. 
The Lagrangian density (\ref{A11}) has the BRS invariance and also BRS exact and thereofere the 
system described by the Lagrangian density is a topological field theory.  
In order to determine the values of $\lambda_{(i)}$, however, we may need infinite numbers of the initial 
conditions, which might be physically irrelevant and the predictability of the theory could be lost. 
This problem could occur because we have not still understood the quantum gravity but we might 
expect that the Lagrangian density (\ref{A11}) might give any clue  for the quantum gravity. 

\section{Summary}

In summary, we have proposed  covariant models which 
may solve one of the problems in the cosmological constant. 
One of the moel can be regarded with an extension of the sequestering model. 
Other modesl can be obtained as an extension of the unimodular gravity in the covariant formulation. 
The contributions to the vacuum energy from the quantum corrections from the matters 
are absorbed into a redefinition of a scalar field and the quantum corrections 
become irrelevant to the dynamics. 
We have also considered some extensions of this model, and we may construct models which may 
describe the realistic evolution of the universe (\ref{NewCCC}). 
We also investigate the properties of the topological models which appear as simple models. 

In this paper, we have not discussed why the vaccum energy is so small but does not vanish. 
In the models, the effective vacuum energy is finite and does not vanish in general. 
We cannot, however, explain why the vacuum energy observed in the present 
universe is so small. 
In our formulation, the problem of the vacuum energy could be reduced from the 
quantum problem to the classical  problem, that is, the problem of the initial condition or the 
boundary condition. 
Then if we find a natural initial condition, say by using the modified gravities, the problem of 
the smallness in the cosmological constant might be solved. 

\section*{Acknowledgments}

The author is indebted to Kugo and Katsuragawa for useful discussions. 
This work is supported (in part) by
MEXT KAKENHI Grant-in-Aid for Scientific Research on Innovative Areas ``Cosmic
Acceleration''  (No. 15H05890) and the JSPS Grant-in-Aid for Scientific 
Research (C) \# 23540296.

\end{document}